%
%
\input harvmac

\overfullrule=0pt
\parindent=0pt

\def\AdSS5{$AdS_5$}
\def\AdS5s5{$AdS_5 \times S^5$}
\def\gy{g_{YM} }

\def\gs{g_{st}}
\def\NSNS{{$NS\otimes NS$}}
\def\RR{{$R\otimes R$}}

\def\calN{{\cal N}}
\def\det{\hbox{\rm det}}

\def\G(#1){\Gamma(#1)}
\def\alphaprime{\alpha'}

\def\C|#1{{\cal #1}     }
\def\(#1#2){(\zeta_#1\cdot\zeta_#2)}


\def\xxx#1 {{hep-th/#1}}
\def\lr { \lref}
\def\npb#1(#2)#3 { Nucl. Phys. {\bf B#1} (#2) #3 }
\def\rep#1(#2)#3 { Phys. Rept.{\bf #1} (#2) #3 }
\def\plb#1(#2)#3{Phys. Lett. {\bf #1B} (#2) #3}
\def\prl#1(#2)#3{Phys. Rev. Lett.{\bf #1} (#2) #3}
\def\physrev#1(#2)#3{Phys. Rev. {\bf D#1} (#2) #3}
\def\ap#1(#2)#3{Ann. Phys. {\bf #1} (#2) #3}
\def\rmp#1(#2)#3{Rev. Mod. Phys. {\bf #1} (#2) #3}
\def\cmp#1(#2)#3{Comm. Math. Phys. {\bf #1} (#2) #3}
\def\mpl#1(#2)#3{Mod. Phys. Lett. {\bf #1} (#2) #3}
\def\ijmp#1(#2)#3{Int. J. Mod. Phys. {\bf A#1} (#2) #3}

\def\Tr{{\rm Tr}}

\def\lam16{\lambda^{16}}

\parindent 25pt
\overfullrule=0pt
\tolerance=10000


\lr\stromingerb{A. Strominger, {\it Loop corrections to the universal hypermultiplet},
hep-th/9706195.}\lr\ferrarab{I. Antoniadis, S. Ferrara, R. Minasian and K.S. Narain,
{\it $R^4$ couplings in M and type II theories},
hep-th/9707013.}
\lr\knn{J.Koplik, A.Neveu, S.Nussinov,  {\it Some aspects of the planar perturbation series}, Nucl. Phys.  {\bf B123} (1977)  109.}
\lr\ew{E.Witten,  {\it Current algebra theorems for the U(1) \lq Goldstone boson'}, Nucl. Phys.  {\bf B156}  (1979)  269;  {\it Instantons, the quark model and the $1/N$ expansion}, Nucl. Phys. {\bf B149}
(1979) 285.}
\lr\jr{R.Jackiw, C.Rebbi, Phys. Rev. {\bf D14}, (1976), 517.}
\lr\hh{G.Horowitz, H.Ooguri, hep-th/9802116 .}
\lr\kslvz{S.Kachru, E.Silverstein, {\it 4-D conformal Theories and Strings on Orbifolds}, hep-th/9802183;
A.Lawrence, N.Nekrasov, C.Vafa, {\it On Conformal Field Theories in Four Dimensions}, hep-th/9803015. }
\lr\thorn{C.B.  Thorn, in Sakharov Conference on Physics, Moscow, (91),447.}
\lr\dpsgketc{S.Gubser, I.Klebanov, A.Peet, Phys. Rev. {\bf D54}, (1996), 3915, hep-th/9602135 ;
 A.A.Tseytlin, I.Klebanov, Nucl. Phys. {\bf B475}, (1996), 179, hep-th/9604166 ;
 S.Gubser, I.Klebanov, Phys. Lett. {\bf B413}, (1997), 41, hep-th/9708005; and references cited in \juan.}
\lr\juan{J.  Maldacena, {\it   The large $N$ limit of  
superconformal field
theories and supergravity}, hep-th/971120.}
\lr\bvz{M.Bershadsky, Z.Kakushadze, C.Vafa, {\it String Expansion as Large N Expansion of Gauge Theories}, hep-th/9803076;
M.Bershadsky, A.Johansen, {\it Large N Limit of Orbifold Field Theories}, hep-th/9803249.}
\lr\gk{S.S. Gubser and I.R.  Klebanov, {\it Absorption by branes  
and Schwinger
terms in the world volume theory}, hep-th/9708005.}
\lr\gkp{S.S. Gubser,  I.R.  Klebanov and A.M.  Polyakov, {\it  
Gauge theory
correlators from non-critical string theory},  hep-th/9802109.}
\lr\wittone{E.  Witten, {\it Anti de Sitter Space and Holography},  
hep-th/9802150.}
\lr\tHS{G. 't Hooft, {\it Dimensional Reduction in Quantum Gravity}, gr-qc/9310006; L.Susskind, {\it The World as a Hologram},
J. Math. Phys. 36, 6377 (1995), hep-th/9409089.}
\lr\greengut{M.B.~Green,  M.~Gutperle and H.~Kwon, {\it Sixteen fermion
 and related
terms in M theory on $T^2$}, \xxx9710151.} 
\lr\nilsson{B.E.W. Nilsson and A.  Tollsten, {Supersymmetrization of
$\zeta(3) R_{\mu\nu\rho\sigma}^4$ in superstring theories}, Phys. Lett
{\bf 181B} (1986) 63.} 

\lr\ggprop{M.B.~Green and M.~Gutperle, {\it Effects of D-instantons}, 
\xxx9701093, \npb498(1997)195.}
\lr\greenschwarz{M.B.  Green and J.H. Schwarz, {\it Supersymmetric
dual string theory (II).  Loops and renormalization}, Nucl.  Phys.
{\bf B198} (1982) 441.}
\lr\osborn{  J.  Erdmenger and H.  Osborn, {\it
Conformally covariant differential operators: Symmetric tensor
fields}, gr-qc/9708040; Class. Quantum Grav. {\bf 15} (1998) 273.} 
\lr\eguchi{T.  Eguchi, {\it S Duality and Strong Coupling Behavior of Large N Gauge Theories with N=4 Supersymmetry}, hep-th/9804037.}
\lr\howest{P.S.  Howe and P.C. West, {\it The complete $N=2$ $D=10$
supergravity}, Nucl.  Phys.  {\bf B238} (1984) 181.}
\lr\matrixth{T.~Banks, W.~Fischler, S.H.~Shenker and  L.~Susskind,  
{\it M
Theory As A Matrix Model: A Conjecture}, \physrev55(1997)5112,  
\xxx9610043. }
\lr\gw{D.J.~Gross and E.~Witten, {\it Superstring modifications of
Einstein's
    equations}, \npb277(1986)1.} 
\lr\gris{M.T.~Grisaru , A.E.M~Van de Ven and D.~Zanon, {\it
Two-dimensional
supersymmetric sigma models on Ricci flat Kahler manifolds are not
finite},
\npb277(1986)388 ; {\it Four loop divergences for the N=1 supersymmetric
nonlinear sigma model in two-dimensions}, \npb277(1986)409.}
\lr\jackreb{R.  Jackiw and C.  Rebbi, {\it Spinor analysis of
Yang--Mills theory}, Phys.  Rev.  {\bf D16} (1977) 1052.}
\lr\greengutc{M.B.~Green and M.~Gutperle, {\it D-particle bound states and the 
D-instanton measure}, \xxx9711107,$\;$  {\bf JHEP01}(1998)005.}
\lr\f{D.Z.Freedman, S.D.Mathur, A.Matusis, L.Rastelli, {\it Correlation
Functions in the CFT(D)/ADS(D+1) Correspondence}, hep-th/9804058 .}  


\noblackbox
\baselineskip 14pt plus 2pt minus 2pt
\Title{\vbox{\baselineskip12pt
\hbox{hep-th/9804170}
\hbox{RU-98-17}
\hbox{DAMTP-98-25}
\hbox{NSF-ITP-98-054}
}}
{\vbox{
\centerline{  Non-perturbative Effects in $AdS_5 \times S^5$ String}
\smallskip
\centerline{Theory and $d=4$  SUSY  Yang--Mills }
  }}
\centerline{Tom Banks\foot{Dept. of Physics and Astronomy, Rutgers University, 
Piscataway, NJ 08855, USA;\ {\it banks@physics.rutgers.edu}}  
and Michael B.  Green\foot{DAMTP, Silver Street, Cambridge CB3 9EW, UK;\ {\it M.B.Green@damtp.cam.ac.uk}}}
\medskip
\centerline{Institute for Theoretical Physics, Santa Barbara, CA   
93106-4030, USA}
\bigskip
 \medskip

\centerline{{\bf Abstract}}
We show that five-dimensional anti de-Sitter space remains a solution
to low-energy type IIB  supergravity when the leading  
higher-derivative corrections to the classical supergravity 
  (which are non-perturbative in the
string coupling)  are included. 
Furthermore, at this order in the low energy expansion of the IIB theory  the 
graviton two-point and three-point functions in $AdS_5 \times S^5$
are shown not to be renormalized 
and  a precise expression is obtained for the four-graviton and related 
S-matrix elements.  By invoking Maldacena's conjectured connection
between IIB superstring theory and  supersymmetric Yang--Mills theory
corresponding statements are obtained concerning correlation  
functions of the
energy-momentum
tensor and related operators in  the large-$N$ Yang--Mills theory.  This leads to interesting
non-perturbative statements and insights into the r\^ole of  
instantons in the  gauge theory.

 \Date{April 1998}

\vfill\eject

\noblackbox
\baselineskip 14pt plus 2pt minus 2pt

\newsec{\bf Introduction}

Recent results on D-brane black hole physics \dpsgketc\ have led to  
a very
interesting conjecture by Maldacena \juan\ which proposes an
exact correspondence between string theory on asymptotically  
Anti-de-Sitter
(AdS) spaces, and certain quantum field theories living on the  
boundary of the
AdS space.  Although the direct
evidence for this conjecture is sparse (it consists primarily of the
identity of multiplicities of short representations of the AdS supergroup
which are found in the two pictures)   it leads to a beautifully  
consistent
picture of the possible  behavior of strongly coupled gauge  
theories\foot{
Furthermore,
an extension of the conjecture
in  \kslvz\ has led to the proof of a striking new result in large-N
gauge theory \bvz.}.  This correspondence also explains
the extrapolation of many D-brane black hole calculations beyond their
apparent range of validity.  One is therefore tempted to simply  
believe the
conjecture and explore its implications for the properties of gauge  
theory
and string theory.

In this paper we will use the Maldacena conjecture to explore  
nonperturbative
properties of
four-dimensional maximally supersymmetric ($\calN =4$) $SU(N)$  
Yang--Mills
theory in the large-$N$ limit.
This theory is related by the  
conjecture to
type IIB superstring theory
compactified on
\AdS5s5.
  This maximally SUSY background is known to be a solution of  
classical type
IIB supergravity
and therefore it is a solution to low energy IIB
 superstring theory to lowest order in the inverse string tension,
$\alphaprime$.
We will exploit knowledge of certain nonperturbative terms in the IIB
effective
action \ggprop\greengut\  to obtain nonperturbative information about the  
large-N gauge
theory.  Of particular interest will be the $R^4$ term \ggprop\
 (where $R$ is the ten-dimensional  Riemann curvature) which arises  
at order
${\alpha'}^3$ relative to the Einstein--Hilbert term.  This term
can be expressed as a particular contraction of four Weyl tensors and  has  a
coefficient $f_4(\rho,\bar \rho)$ which is an exactly known
modular function of the complex scalar field $\rho = c^{(0)} + i  
e^{-\phi}$,
where
$\phi$ is the type IIB dilaton (so that the string coupling is $\gs  
= e^\phi$)
and $c^{(0)}$ is the Ramond--Ramond (\RR)
scalar.  Other interaction terms of the same dimension are also  
known. They
are
related to
this term by supersymmetry.

According to Maldacena's conjecture, which we
review
below, the region of validity of the $\alphaprime $ expansion  
translates in
gauge
theory into the region of  large $\gy^2 N$, where $\gy$ is the  
Yang--Mills
coupling constant.  One way of achieving this is to hold $\gy$ fixed
and take $N\to \infty$, which is of relevance to the Matrix approach
\matrixth. In the 't Hooft limit   $\gy^2 N = \hat g^2$ is fixed with
$N\to \infty$ so that $\gy^2 =\hat g^2/N \to 0$.  
The $\alphaprime$ expansion is relevant in this limit
only for strong coupling (large values of $\hat g$). The common domain of validity of the
two expansions is that of low energy, perturbative, string theory.
On the other hand, since we have nonperturbative information
about string theory we can explore the gauge theory when $N$ is 
large, even outside the range of validity of the 't Hooft expansion.
  The nonperturbative terms in string theory which we discuss,
lead to  contributions to  
gauge theory
correlators which are  of a particular order in the strong coupling  
expansion
in powers of $(\gy^2 N)^{-\ha}$, but nonperturbative in $\gy$ itself.

The first application of our results will be  to prove that the
\AdS5s5\ solution is an exact solution of string theory when the  
$O(\alphaprime^3)$ nonperturbative terms are included.
This small step towards
proving that it is a consistent background for the full string theory
follows essentially from the fact that \AdS5s5\ is conformally flat.
Furthermore, our results are in accord with  certain nonrenormalization
theorems for two-point
functions proven directly
in the gauge theory \gk\ and suggest a new one for three-point functions\foot
{Although the existence of three-point function nonrenormalization theorems is
known to some experts, there is no systematic discussion of them
in the literature.  A recent paper \f\ provides a proof of a subset
of these theorems.}.
These theorems are the direct analog of the familiar  non-renormalization
theorems
for graviton two-point and three-point functions in ten-dimensional  
flat space  superstring
theory, here
generalized (at $O(\alphaprime^3)$) to the \AdS5s5\ background.   
The $R^4$ term
in IIB supergravity translates into an
exact (in $\gy$) formula for the connected four point function of
stress tensors to next to leading order in the strong coupling expansion.
We will see that the $SL(2,Z)$ duality symmetry of the IIB string theory
translates into a precise statement concerning the way in which the
corresponding modular transformations of  $\calN=4$ Yang--Mills act  
on these
correlation functions in the large-N limit.  Related statements  
will be deduced
for other correlation functions, notably the correlation of  
sixteen
fermionic spin-half superpartners of the lagrangian density.  Since the
function
$f_4$ (and its
supersymmetric relatives) can be expanded for small $\gs$   as an  
exact sum
over D-instanton contributions in the string theory we are able to  
make precise
statements about Yang--Mills instanton contributions to the gauge  
theory for
certain classes of correlation functions in the  small $\gy$ limit  
(with $\gy^2
N$ fixed and large).

\newsec{\bf The Type IIB --- SYM Dictionary}

According to the interpretation of \juan\ given in \gkp\wittone\
there is a precise mapping of the S-matrix elements of ten-dimensional
superstring theory in the \AdS5s5\ background\foot{
We will use the term S-matrix elements to refer to
the effective action as a function of boundary values
described by \gkp\ and \wittone, even though wave packets
in AdS space do not separate and the usual concept of an S-matrix
is ambiguous.  These are certainly limits of S-matrix elements in
the asymptotically flat space which exist before the Maldacena
limit is taken.  We further believe that in perturbative string theory,
these amplitudes will be expressed as expectation values of BRST
invariant vertex operators in the \AdS5s5 $\sigma$ model.  We thank
E.Witten and D.Freedman for discussions of this issue.}
 to correlation functions of
operators in the large-$N$ Yang--Mills theory which lives on the  
boundary of
\AdSS5. One starts by solving the field equations deduced from  the  
effective
action for the string theory, $S[\Phi]$,  with specified boundary  
conditions
for
the fields on the four-dimensional  boundary  of \AdSS5, $\Phi|_{\partial
( AdS_5)}=
\tilde \Phi$.  The expression $S[\tilde \Phi]$ is then interpreted as the
generating functional for correlation functions of operators in the
four-dimensional   superconformal Yang--Mills theory living on  the  
boundary.
Thus, a correlation function of $K$ operators has the form,
\eqn\corrdef{{\delta \over \delta \tilde \Phi_1} \dots   {\delta  
\over \delta
\tilde \Phi_K}  S[\tilde \Phi],}
where the operators are located at points $y_1,\dots , y_K$ in the
four-dimensional boundary.  It is easy to see that this definition  
identifies
the correlation functions with a quite precise 
analog of the $K$-particle S-matrix element in the
superstring theory. That is, both the S-matrix of
string theory in Minkowski space, and the objects studied in
\gkp\ and \wittone\ are obtained by solving the effective
equations of motion with boundary conditions at infinity.
Note that in this analogy, ``the incoming and outgoing on-shell
states\rq\rq{}
are localized
at asymptotic points labelled by $y_k$, rather than being momentum
eigenfunctions.
 The on-shell condition  
determines the
behaviour of the field as a function of the extra \AdSS5\  
coordinate, $U$.

This identification emphasizes the holographic \tHS\ nature of  
string theory
and shows us that it is related to the well known fact that the  
only physical
quantities
in the theory are on shell S-matrix elements.  In field theory, the  
S-matrix is
computed in
terms of the generating functional with sources which are  
nonvanishing only at
infinity.
The existence of a gauge invariant off shell continuation of  
Green's functions
is intimately related to
locality of the underlying theory: field theory is not holographic.
Conversely, the holographic nature of string theory is implicit
in the statement that only the on shell S-matrix is observable.
This statement of the holographic principle in string theory is  
more general
and more covariant
than the Thorn-Susskind \thorn\tHS\ \lq\lq wee parton\rq\rq{}
ansatz in
the light cone gauge.

The correspondence between couplings in the Yang--Mills  and string theories  
is given
by
\eqn\gym{\gy^2 =4\pi \gs, \qquad \theta =2\pi c^{(0)},}
so that 
\eqn\sdef{S \equiv { \theta\over 2\pi} +{ 4\pi i\over  \gy^2  } =
\rho,}
and the combination $\gy^2 N$ is given in terms of the string  
theory parameters
by
\eqn\gyn{\gy^2 N = L^4 \alphaprime^{-2} ,}
where $L$ is the radius of curvature of the $AdS_5$ space.  The 32  
components
of
the local supersymmetry charges in the IIB superstring translate  
into the 32
components of the rigid  superconformal symmetry of the $\calN =4$  
$SU(N)$
Yang--Mills theory.   Although Maldacena's conjecture is supposed  
to be valid
for any values of $\gy^2$ and $N$, one can only do computations in certain
limits.
As is typical in dual situations, the regions of validity of gauge  
theory and
string theory
computations are complementary.   Only quantities protected by
nonrenormalization
theorems can be easily computed in both languages.  However, if  
we believe
the conjecture,
it immediately tells us many things about both theories.  For  
example, we are
led to believe
that the gauge theory has a large $N$ limit for any value of $\gy$,  
not just in
the 't Hooft
regime.    Weakly coupled string theory explores the large coupling  
limit of
the planar gauge
theory (large $\gy^2 N$), while the $\alphaprime$ expansion at arbitrary string  
coupling explores
the large $N$
limit at arbitrary Yang--Mills coupling, $\gy$.  Perturbative gauge theory  
computations (small $\gy^2 N$) are
valid only in the regime
where string theory lives on a space-time of sub-stringy scale.

\newsec{\bf $R^4$ Terms in IIB String Theory on \AdS5s5}

The part of the low-energy effective IIB supergravity action which  
will concern
us has the form (in string frame)
\eqn\iibact{S^{IIB} = {1\over \alphaprime^4} \int d^{10} x \sqrt{-G}
(e^{-2\phi} R + k {\alpha'}^3 e^{-\phi/2} f_4(\rho,\bar \rho) R^4
+ \cdots),}
where $k$ is a known constant and we have indicated by
$\cdots$ the terms which depend on any field other than the metric  
and the
complex scalar $\rho$.  The fact that there is no known action for the
self-dual five-form field
strength, $F_5$,  will not concern us since we will only make use of its
equations of motion, which are solved, in the  \AdSS5\  background, by   
$F_5 = c
\epsilon_{\mu_1 \dots \mu_5}$, where the constant $c$ determines the
cosmological constant.

The $R^4$ term in \iibact\ has a coefficient $f_4(\rho,\bar \rho)$  
which is a
non-holomorphic modular function of the scalar field as required by the
$SL(2,Z)$ symmetry of the type IIB theory  \ggprop.  It is given by the  
nonholomorphic
Eisenstein series,
\eqn\fdef{ f_4(\rho,\bar \rho) = \sum_{(m,n) \ne (0,0)}   {\rho_2^{3/2} \over |m+  
\rho n|^3},}
which may be expanded for large $e^{-\phi}$ (small string coupling) as
\eqn\fexpan{\eqalign{& e^{-\phi/2}  f_4 \sim  2 \zeta(3) e^{-2\phi} +  
{2\pi^2 \over
3}
\cr
&  +(4\pi)^{3/2} e^{-\phi/2}  \sum_{M >0} Z_M M^{1/2}
\left(e^{-2\pi M (e^{-\phi} + i c^{(0)})} + e^{-2\pi M (e^{-\phi} - i c^{(0)})} \right)  
\left(1 +
o (  e^{\phi} /M )   \right).\cr}}
The first two terms of this expansion are the tree-level and one-loop
contributions  of string perturbation theory while the remaining  
terms are a
sum
over charge-$M$  D-instanton contributions, each of which has an  
infinite power
series of perturbative corrections (which are explicitly given in
\ggprop). The coefficient $Z_M$ is given by
\eqn\zmdef{Z_M = \sum_{m|M} {1\over m^2},}
where $m|M$ denotes that the sum is over the divisors of $M$.

 The symbol $R^4$ is  used in \iibact\ as a shorthand way of writing a
particular contraction of four Riemann tensors which can be conveniently
expressed as an integral over an auxiliary Grassmann variable $\theta$ which is a
sixteen-component complex chiral $SO(9,1)$ spinor  \nilsson\ggprop,
\eqn\rfourdef{R^4  = \int d^{16}\theta (R_{\theta^4})^4,}
where
\eqn\rthetdef{R_{\theta^4} \equiv \bar \theta
\gamma^{\mu\nu\sigma} \theta \bar \theta \gamma^{\rho\tau}_{\ \  
\sigma} \theta
\,
R_{\mu\nu\rho\tau}.}
Here $\bar \theta = \gamma^0 \theta$ and the ten-dimensional gamma  
matrices
with world indices are defined by
\eqn\gamdef{ \gamma^\mu = e_a^{\ \mu} \gamma^a,}
where $\gamma^a$ are the usual $SO(9,1)$ gamma matrices, $e_\mu^{\ a}$
($m=1,2,3,4$)  is  the
zehnbein  and $a$ is a ten-dimensional tangent-space index ($\mu,a  
= 0, 1,
\cdots, 9$).  Equation \rfourdef\  implies that $R^4$ transforms as a
scalar
density  under  general coordinate transformations.

We now wish to consider the compactification of the theory on  
\AdS5s5, which
is
known to be a solution to the classical low energy theory with a
non-vanishing $F_5$.  It is of great significance that the only components of the curvature
which contribute in the expression $R_{\theta^4}$ are those of the Weyl
tensor, $C_{\mu\nu\rho\tau}$.  Since the \AdS5s5\ is conformally flat
it follows that the $R^4$ interaction vanishes in this background.  At
the risk of seeming pedantic, we will here  demonstrate this explicitly.
 The curvature for this background (which is the  product of two symmetric  
spaces) breaks
up into two disjoint pieces associated with different directions.   
We shall use
a tangent-space basis for the curvature (so that all the zehnbeins  
will drop
out
of \rfourdef) in which the \AdSS5\ directions $a,b,c,d = 0,1,2,3,4$ are
labelled
$M,N,P,Q$ so that,
\eqn\rdefone{R_{abcd} \equiv R^{(1)}_{MNPQ} = -L^{-6} (\eta_{MP} \eta_{NQ} -
\eta_{MQ} \eta_{NP}),}
where $L$ is the \AdSS5\ length scale and $\eta_{MP}$ is the signature $(4,1)$ Minkowski metric. The components in the $S^5$  
directions
$a,b,c,d = 5,6,7,8,9$ are labelled $m,n,p,q$ and are given by,
\eqn\rdeftqo{R_{abcd} = R^{(2)}_{mnpq} =  L^{-6} (\delta_{mp} \delta_{nq}  
- - \delta_{mq}
\delta_{np}),}
and all other components of the curvature vanish.

Substituting these expressions into \rthetdef\ gives
\eqn\rthet{L^6 R^{(0)}_{\theta^4} = \bar \theta \gamma^{ABC}\theta\, \bar
\theta
\gamma_{ABC}\theta -  \bar \theta \gamma^{abc}\theta\, \bar \theta
\gamma_{abc}\theta + \bar \theta \gamma^{AB}\gamma^{c}\theta\, \bar  
\theta
\gamma_{AB}\gamma^c\theta - \bar \theta  
\gamma^{A}\gamma^{bc}\theta\, \bar
\theta \gamma_{A}\gamma^{bc}\theta.}
This expression has a manifest symmetry under the $SO(4,1) \times SO(5)$
subgroup of $SO(9,1)$.  In this decomposition the $SO(9,1)$ spinor  
is written
as
a bi-spinor in the $(4,4)$ representation so that $\theta \equiv
\theta_{\alpha_1, \alpha_2}$ where the subscripts 1 and 2 label the  
$SO(4,1)$
spinor and the $SO(5)$ spinor, respectively.
  The  symbol $\bar \theta$  is defined
similarly for each of the  two groups. For  a $SO(4,1)$ spinor $\bar
\phi_{\alpha_1} = (  \phi\gamma^0_1)_{\alpha_1}$.  For   $SO(5)$  
(which is
pseudoreal) the bar is defined by  $\bar \phi_{\alpha_2}
=  (\phi^* J)_{\alpha_2}$ with $J J^\dagger  =1$.  The similarity  
between the
two  groups is exhibited most clearly if a Majorana representation  
is chosen
for
$SO(4,1)$ in which all the gamma matrices, $\gamma^M_1$, are real  
and the same
form is chosen for the gamma matrices, $\gamma^m_2$ ($m =1,\cdots,4$), of $SO(5)$ but with
$\gamma^0_2 = i \gamma^0_1$.  In this basis the  hermitian matrix    
 $J$ is
given by $J= i\gamma^0_2$.  We therefore conclude that in this basis,
\eqn\bardef{\bar \theta = i\theta^* \gamma^0_1 \gamma^0_2.}
Substituting this into \rthet\ it follows immediately that
\eqn\mainres{R^{(0)}_{\theta^4} =0,}
 since the difference in signature between $SO(4,1)$ and $SO(5)$ does not
affect
the contractions in  \rthet.  This, of course, is a particular
property of the compactification on \AdS5s5\  which  follows from the
fact that it is conformally flat (so the Weyl tensor vanishes) as
mentioned earlier.  This is not a property of more general backgrounds\foot{For example, compactification on Calabi--Yau threefolds has been considered in \stromingerb\ferrarab.}.

This very  simple result leads immediately to several important
consequences.  We see from \rfourdef\ that  $R^4 =0$ in the \AdS5s5\
background.  This means that the dilaton equation of motion, which  
has a term
proportional to $R^4$, is  unchanged in this background.  We also  
see that
$\delta R^4 / \delta g_{\mu\nu} =0$ (since the differential is
proportional to $(R_{\theta^4}^{(0)})^3$), which means that the   
Einstein
equation is also unaffected.  It is also easy to see that none of  
the other
$O(\alphaprime^3)$ terms, which are related to $R^4$ by supersymmetry,
 contribute
to the equations of motion. This means that the \AdS5s5\ background is
unaltered
by the presence of this term, so this background is a solution of  
the effective
equations of motion of string theory through $o(\alphaprime^3 )$,  
to all orders
in $g$.  As yet there is no $\sigma$-model argument which  
shows that
the \AdS5s5\ background is a solution of tree level string theory
although this is undoubtedly true.  Such
arguments
usually depend on world sheet superconformal invariance which is  
broken by
the five form background.

 Equation \mainres\ also implies that
 \eqn\twothree{{\delta \over \delta g_{\mu\nu}} {\delta \over \delta
g_{\rho\sigma}} R^4 \big|_{AdS_5\times S^5} =0 =    {\delta \over \delta
g_{\mu\nu}} {\delta \over \delta g_{\rho\sigma}} {\delta \over \delta
g_{\tau\omega}} R^4 \big|_{AdS_5\times S^5}.}
This shows that there is no renormalization of the graviton two-point or
three-point functions in the \AdS5s5\ background at  
$O(\alphaprime^3)$. These
are expected to extend to exact non-renormalization theorems  
analogous to those
which
 are known to
be true in string perturbation theory around ten-dimensional  
Minkowski space.

 There is a non-zero four-graviton contribution from the $R^4$ term  
which is
obtained by differentiating  four times with respect to the metric, which
leaves
no overall powers of $R_{\theta^4}^{(0)}$.  This adds to the classical
term which arises from the Einstein--Hilbert action so that up to
$O(\alphaprime^3$) the four-graviton amplitude is proportional to   
\eqn\fourst{e^{-2\phi} A_4^{String\,
(1)} +k \alphaprime^3 e^{-\phi/2} f(\rho,\bar \rho)  A_4^{String \,(2)},}
where $A_4^{String\, (1)}$  is the classical
amplitude obtained from the Einstein--Hilbert action while
$A_4^{String\, (2)}$ is the contribution from the $R^4$ term. 
The new term 
can be computed in terms of the four-point vertex in the effective
action, which is proportional to
\eqn\vertex{S_4 =\alphaprime^3 \int d^{10}x \sqrt G e^{-\phi/2} f(\rho, \bar \rho)
 t_8^{\mu_1\omega_1 \dots \mu_4 \omega_4}
t_8^{\nu_1\tau_1 \dots \nu_4\tau_4}
R_{\mu_1\omega_1\nu_1\tau_1 } \dots
R_{\mu_4 \omega_4\nu_4\tau_4}.}
The scalar field $\rho$ is set to its constant background value while the  linearized curvature is
\eqn\lincurv{R_{\mu_1\omega_1\nu_1\tau_1} = D_{\omega_1} D_{\tau_1}
h_{\mu_1\nu_1},}
where  $h_{\mu\nu}$ is the linearized fluctuation of the metric around
its value in   \AdS5s5\ and $D$ is the \AdS5s5\ covariant derivative.  
The symmetries of
$R_{\mu\omega\nu\tau}$ are imposed by the symmetries of the tensor $t_8$
(defined by eq. (2.16) of \greenschwarz)  which has the form of  the product of four
inverse \AdS5s5\ metrics summed over various permutations
 of their indices.   The boundary Yang--Mills field theory that we 
are interested in  will be obtained by substituting the solution of
Einstein's equations  linearized
around \AdS5s5\  for $h_{\mu_r\nu_r}$, 
with the boundary condition that it approaches the Minkowski plane wave
with specified  momenta, at infinity. In particular, correlations of
the Yang--Mills stress tensor will arise from  
the components of $h_{\mu_r\nu_r}$ which are oriented
 in the four-dimensional Minkowski directions, $\mu_r = M_r$, $\nu_r =
N_r$ where $M_r,N_r = 0,1,2,3$.  It is important that it is the
ten-dimensional momenta which satisfy an on-shell constraint and not the
four-dimensional
Minkowski momenta, $k_{M_r}$.  If we restrict
consideration to s-waves with respect to the $S^5$ then only  the $AdS_5$ 
part of the ten-dimensional metric is important and this has the form,
\eqn\metric{ds^2 = U^2 (- dt^2 + (d{\bf x})^2 ) + {dU^2 \over U^2}}
where ${\bf x}= \{x_0,x_1,x_2,x_3\}$ and $U= x_4$. 
The on-shell condition determines the $U$ dependence of $h_{\mu\nu}$
in terms of the Minkowski momenta.   From these expressions we can, by
Fourier transformation, obtain the objects which are supposed to
match with local correlation functions of stress tensors
 in the supersymmetric Yang--Mills  theory.

\newsec{\bf Non-perturbative Terms in Large-$N$ SUSY Yang--Mills Correlation
Functions}

Using the Maldacena conjecture, our results for the string  
effective action in
\AdS5s5\ may be converted into statements about correlation  
functions in SYM
theory.
We will concentrate mainly on statements about the scattering of gravitons
with polarizations in the \lq\lq Minkowski\rq\rq{} directions of $AdS_5$,
which 
translate into properties of  correlation functions of the SYM  
stress tensor,  $T_{\mu\nu}$.
Many other   correlation functions
 are related to these by
supersymmetry.
The coefficient of the $R^4$ term in \iibact\ has the prefactor  
$(e^{\phi/2}
\alpha')^{3} f(\rho) = L^6 N^{-3/2} f(S)$ relative to the  
Einstein--Hilbert
term, which means that it is a non-leading term in the $1/N$ expansion.

The vanishing of the graviton one-point function, is simply the  
statement that
the one-point function of the stress tensor $\langle T_{\mu\nu} \rangle$ vanishes, which follows from
conformal
invariance.  Similarly, the vanishing of the $R^4$ contribution to the
two-graviton S-matrix element in IIB supergravity translates into  
the statement
that the  correlation function of two stress tensors in $\calN = 4$  
Yang--Mills
theory, $\langle T_{\mu_1\nu_1}(1)T_{\mu_2\nu_2}(2) \rangle$, is given by its free field value. This is known to be an  
exact statement
 \osborn\gk\ by virtue of  the relation of this correlation  
function to the
$R$-symmetry anomaly, which is not renormalized due to the   
Adler--Bardeen
theorem.   An analogous Ward identity prevents the three-point  
correlation
function of the stress tensor, $\langle T_{\mu_1\nu_1}(1) T_{\mu_2 \nu_2}(2) T_{\mu_3\nu_3}(3) \rangle$ from receiving renormalizations  
beyond those of
the free field theory\foot{We are grateful to Hugh Osborn for  
explaining this
to us.} which is in accord with the fact that the three-graviton  
amplitude in
 string theory is not renormalized from its classical value.  
It is  
only when we
come to the four-graviton amplitude that the $R^4$ term contributes and
therefore the correlation  function of four stress tensors gets a new
contribution.

{}From the supergravity calculations in the last section we obtain the
following
expression for the momentum-space  correlation function of four  
stress tensors
in the Yang--Mills theory,
\eqn\fourtens{A_4^{YM}= \langle T_{\mu_1\nu_1}(1)T_{\mu_2\nu_2}(2) T_{\mu_3\nu_3}(3) T_{\mu_4\nu_4}(4) \rangle  = A_4^{(1)} + \tilde k N^{-3/2} f(S) A_4^{(2)} + \dots ,}
where  $A^{(1)}_4$ is the contribution  at leading order in  
$(\gy^2N)$
and $A_4^{(2)}$ is  the correction arising from the $R^4$ term (and an
irrelevant constant has been absorbed into $\tilde k$).  
In writing these equations we are using the correspondence between field theory correlators
and string theory S-matrix elements.  In particular, since we are using the effective action,
we need only compute tree diagrams.  The term $A_4^{(1)}$ comes from  the four-graviton amplitude
computed from the Einstein action in \AdS5s5, while $ A_4^{(2)}$ corresponds to the  four-graviton vertex in the $R^4 $ term.  No further terms are necessary because we 
are working to next to leading order in the $\alphaprime$ expansion.  The asymptotic states in both terms
are taken to be delta functions in the boundary Minkowski space.   
The supergravity field
$\rho$ has here been  interpreted as the complex coupling constant, $S=  
\theta/2\pi  +
4\pi i/\gy^2$.  In writing \fourtens\ an overall factor of  
$\alphaprime^{-4} =
(\gy^2
N)^2$  has been absorbed  into the normalization of the stress  
tensors.    In string theory scattering amplitudes, we are using the normalization in which all tree level
amplitudes are of order one, while in the gauge theory we use the corresponding normalization in which all
connected correlation functions of single trace operators are of order one in the large $N$ limit.
Terms of higher order in $(\gy^2 N)^{-\ha}$ can be  
neglected if
$1 >>
\alphaprime^{-2}L^4 = (\gy^2 N)^{-1}$.

The second term in  \fourtens\ has a remarkable amount of information
concerning the Yang--Mills theory.  It is the first non-leading  
term in the
$1/N$ expansion but is an exactly known function of the (complex)  
coupling.
The factor $f(\rho)$, defined in \fdef, has the expansion \fexpan\  for  small $\gs$ ($\rho \to  
\infty$) which
starts with the tree-level term $f \sim \gs^{-3/2} =(4\pi)^{3/2} \gy^{-3} $.   
Therefore, in
the limit
$\gy\to 0$ with $\gy^2 N$ fixed and large, the expression  
\fourtens\ takes the
form 
\eqn\fourtenstwo{\eqalign{ &  A_4^{YM}   = A_4^{(1)}  + \tilde k
A_4^{(2)} \left[2\zeta(3) \left({\gy^2 N\over 4\pi}\right)^{-3/2}
  +  {2\pi^2 \over 3 N^2} \left({\gy^2 N\over
4\pi}\right)^{1/2}\right. \cr &\left. +  {(4\pi)^{3/2}\over N^{3/2}}  
\sum_{M=1}^\infty
Z_M  M^{1/2} \left(e^{- M \left(8\pi^2  \gy^{-2}  +  i \theta  
\right)} + e^{-M \left(8\pi^2  \gy^{-2}  - i \theta \right)}\right)(1 + o(\gy^2 /M ))\right]
 ,\cr}}
which includes an  infinite series of instanton corrections.
We will return to a discussion of these corrections in the next section.

It is important to note that S-duality (the Montonen--Olive duality),
which is a discrete group which
maps the small coupling regime to large coupling,
is manifest in \fourtens\ but  not  in \fourtenstwo, which
is the 't Hooft expansion, and  is only valid when $\gy$ is very small.
We remark in passing that our result does not agree with the general form suggested in
a recent paper of Eguchi \eguchi, who obtained constraints on the strong coupling
behavior of the theory   by insisting that duality
be implemented in the 't Hooft expansion.  When translated into string theory language, one of these constraints
implies that the $\alphaprime$ expansion at fixed $g$ contains only even powers of $\alphaprime$.
The Einstein and $R^4$ terms however differ by three powers of 
$\alphaprime$.\foot{We thank Steve Shenker for a discussion of this point.}

\newsec{\bf Other Non-perturbative Contributions --- Instanton Effects}

An easy way of determining all the interactions in the effective IIB
supergravity action which are of the  
same dimension
as $R^4$ and are related to it by supersymmetry is to use an  
on-shell linearized
superspace formalism.  Thus, the chiral superfield $\Phi(\theta)$  
is a function
of the same 16-component chiral Grassmann spinor which we introduced  
earlier.
The chiral constraint is $\bar D\Phi =0$ (where $D$ is the supercovariant
derivative) and the constraints $D^4 \bar \Phi = \bar D^4 \Phi =0$  
eliminate
the auxiliary components of the field \howest, which has an  
expansion in terms
of the on-shell physical fields,
\eqn\expphi{\Phi(\theta) = \rho + \theta \lambda + \bar \theta
\gamma^{\mu\nu\rho}\theta G_{\mu\nu\rho} + \dots + R_{\theta^4} +  
\dots   ,}
where $\lambda$ is the complex spin-1/2 \lq dilatino',  $G$ is a complex
combination of the \RR\ and \NSNS\ field strengths and the dots  
indicate a
series of terms which terminates at the power $\theta^8$.  The (Weyl)
curvature enters in $R_{\theta^4}$ which has four powers of $\theta$.
 The general
interaction of the type we are concerned with is given by an  
integral of the
form $\int d^{16}\theta F[\Phi]$, which selects out terms with  
sixteen powers
of $\theta$.   These are the terms which originate from the  
integration over the
sixteen fermionic zero modes associated with the   
supersymmetries which
are broken by the presence of a D-instanton.   Among these is the  
$R^4$ term  as well as many others.  Notably, there is a
sixteen-fermion term of the form $\int d^{10} x\, \det e\, e^{-\phi/2} f_{16} (\rho\bar\rho) \lambda^{16}$  which was  
considered in
detail in \greengut.  This is the analogue  of the 't Hooft vertex in
Yang--Mills theory. The symbol $\lambda^{16}$ indicates the fully
antisymmetrized product of the sixteen chiral spinor fields and  
the function $f_{16}$ is given in \greengut\ as,
\eqn\fnewdef{\eqalign{f_{16}(\rho,\bar\rho) =&\Gamma(27/2) \rho_2^{3\over 2}\sum_{(m,n) \ne
(0,0)} {(m+ n \bar \rho)^{24} \over |m+ n \rho |^{27}} \cr
 \sim & \Gamma(27/2)\zeta(3)e^{-3\phi/2} + \Gamma(23/2) e^{\phi/2}\cr
&\ \    +
\pi^{-{1\over 2}} (4\pi e^{-\phi})^{12} \sum_{M>0} Z_M M^{25\over 2}
e^{- 2\pi M (e^{-\phi} + i c^{(0)})}(1+ o(e^\phi/M)), \cr}}
where the second line includes only the leading contributions for
small $\gs=e^\phi$  to each
term in the instanton sum (the anti-instantons are suppressed by
powers of $e^{\phi}$).  We see that there are perturbative tree
and one-loop terms in addition to the infinite set of D-instanton terms.
The classical IIB supergravity is invariant under $SL(2,R)$
transformations, under which $\lambda^{16}$ transforms by an arbitrary phase
(assuming we are working with the scalars in the coset
$SL(2,R)/U(1)$).  However, the   expression for $f_{16}$ is  a modular
form with
holomorphic and anti-holomorphic weights $(12,- 12)$ which transforms
with a discrete phase under $SL(2,Z)$, 
\eqn\trans{f_{16} \to \left({c\rho + d\over c \bar \rho +
d}\right)^{12}\, f_{16},}
(integer $c$ and $d$),  thereby
cancelling only the subset of the transformations of $\lambda^{16}$ which
 are in
$SL(2,Z)$.  This reflects the fact that in the string theory the
continuous $SL(2,R)$ is not a symmetry, even at tree level.   Similar non-holomorphic  modular forms are associated with other interactions that are related to $R^4$ by supersymmetry, such as $f_8 G^8$ and many others.

We now turn to the corresponding description of the $\calN=4$ SUSY
Yang--Mills theory.  A  charge $M$ D-instanton contributes a weight $e^{-
2\pi M/\gs}$ to an S-matrix element in the IIB theory.  Using the
dictionary, this  translates
into a contribution to the corresponding process in the Yang--Mills
theory of a contribution with 
weight $e^{-8M\pi^2 /\gy^2}$, which is the contribution of a
charge-$M$ Yang--Mills instanton.  This means that the instanton
contributions to the $R^4$, $\lambda^{16}$ and other IIB  interactions
discussed earlier must have a direct interpretation in large-$N$
Yang--Mills theory.   Intuitively, this is clear by considering the euclidean D3-brane/D-instanton configuration, which preserves half of the $\calN=4$ supersymmetries.  This system has a \lq Higgs' branch in which the D-instanton is represented by a fini

te-sized  Yang--Mills 
instanton in the D3-brane.

This argument for  interpreting  the D-instanton process
as a Yang--Mills instanton effect in the boundary conformal
Yang--Mills theory is further motivated by  counting the  fermionic
zero modes.
We have already seen that in the IIB theory there are sixteen
fermionic zero modes in a D-instanton background which correspond to
the broken  supersymmetry transformations.  
This leads,  to the $R^4$,  $\lambda^{16}$ and other
terms.  In the classical 
$\calN=4$ Yang--Mills theory an instanton background
has $8N$ fermion zero modes, $2N$ for each of the $4$ adjoint Weyl fermions. 
However, our string computation is valid only in the region
of large $g^2 N$.  Most of the classical zero modes are not related
to symmetries.  In particular, there are no discrete remnants of anomalous
$U(1)$ symmetries which dictate the number of fermions in the 't Hooft
interaction.  The superpotential terms (in $\calN = 1$ language)
of the $\calN = 4$ theory break all classical symmetries apart from
the $SU(4)$ R symmetry.   Thus, in planar perturbation theory around the
instanton, the superpotential can convert the classical 't Hooft vertex
into terms with different numbers of external fermion legs.  The only zero
modes which are protected are those which follow from supersymmetry.

Those superconformal generators which fail to annihilate the instanton
lead to normalizable zero modes
\jackreb .  There are precisely sixteen such superconformal zero modes which means that there must be a correlation function of
sixteen spin-1/2 operators in the Yang--Mills theory which have the
$\tilde\lambda$'s as their fermionic sources (where $\tilde \lambda$
is the boundary value of the dilatino).  These operators can be
obtained by a supersymmetry transformation on $\Tr F^2$ and have the
form $\Psi = \gamma^{\mu\nu}\Tr (F_{\mu\nu} \psi) + \psi^3\ {\rm terms}$  (where 
we are using
 ten-dimensional notation to label the $\calN=4$ fields).
Our calculation determines the form of the $\Psi^{16}$ vertex in the
large $g^2 N$ limit.   

As with the correlation of four stress tensors, this correlation
function of sixteen  fermionic operators,  can be extracted from the IIB
expression  by using the dictionary and has the form, 
\eqn\sixteenf{  N^{-3/2} 
f_{16}(S) \Psi^{16}.}
 The invariance of this expression under $SL(2,Z)$
transformations  is again manifest.  The expansion \fnewdef\ can now
 be used to expand this expression for small Yang--Mills coupling $\gy$.  Even  though  the  individual
Yang--Mills instanton terms  are obviously highly suppressed since they have
factors of $e^{-8\pi^2 M\gy^{-2}}$ the full instanton sum is
crucial for ensuring that the correlation functions transform with
appropriate weight under $SL(2,Z)$ S-duality transformations.    

The leading term in the small coupling 
limit in which $\gy^2N$ is fixed and large is again a term of order
$(\gy^2N)^{-3/2}$, which comes from the tree-level term in \fnewdef\
and is not suppressed by powers of $N$.
  Thus, the strong coupling expansion 
of the sum of planar diagrams will give a nonzero contribution to the Green's function
of sixteen powers of $\gamma^{\mu\nu}\Tr (F_{\mu\nu} \psi)$.  Note that
unlike the 't Hooft interaction
in gauge theories with less SUSY, this does not break any classical $U(1)$ symmetry.
Indeed, the superpotential (in $\calN = 1$ language) of the $\calN = 4$ theory
breaks all potentially anomalous symmetries at the classical level.  The full classical
global symmetry is the $SU(4)$ R symmetry and this is not anomalous.   It is 
preserved by the $\Psi^{16} $ term.   

Since the Yang-Mills instantons do not break any symmetry of the gauge theory,
and since the perturbation expansion in $\gy $ is only asymptotic,
one might wonder how one could separate their contributions from 
ambiguities in the resummation of the perturbation series.  The answer 
is again obtained by appealing to string theory.  There it is known
that all amplitudes are independent of the constant mode of the 
\RR\ scalar, to all orders in string perturbation theory. 
In the gauge theory, the constant mode of the scalar is just
the $\theta$ parameter.   String theory therefore predicts that SYM Green 
functions are independent of $\theta$ to all orders of the $1/N$ expansion,
but should pick up $e^{-N}$ contributions which are periodic in $\theta$.

This prediction appears to be verifiable by direct calculation in Yang Mills
theory.   Indeed, it is well known that correlation functions are independent
of $\theta$ in Yang-Mills perturbation theory.  However in purely bosonic
Yang Mills theory, it is expected \ew\ that the sum of all planar diagrams
does depend on $\theta $.   What distinguishes the $\calN = 4$ theory is
its conformal invariance.  The Green functions should be completely
determined in terms of anomalous dimensions and operator product
coefficients.  The perturbative series for these quantities are not 
expected to have renormalon singularities, and therefore, in each
order of the 't Hooft expansion, these series should be convergent \knn.
Consequently, finite orders in the $1/N$ expansion should be independent
of $\theta$ for all $g^2 N$.  $\theta$ dependent corrections should be periodic
and of order $e^{-N}$.  This is precisely the form of the nonperturbative
terms which are predicted by string theory via the Maldacena conjecture.

It would be very interesting to go further 
and make a more precise comparison between D-instanton
and SYM instanton calculations.  In particular, the exact form of the
D-instanton
contributions to the processes described earlier gives a precise
prediction of the measure to be associated with an $M$-instanton
contribution.  However, we should remember that our string computation
only determines the large $\gy^2 N$ behavior of instanton amplitudes.
Thus, we should only expect to match semiclassical Yang Mills
calculations for quantities which obey some sort of nonrenormalization
theorem.

\newsec{\bf Discussion}

By combining nonperturbative information from the low energy IIB string-theory
effective action with the Maldacena conjecture we have arrived at a
number  of perturbative and nonperturbative predictions about the maximally SUSY Yang Mills 
theory in four dimensions.  For example,   it  leads to a simple
explanation of the nonrenormalization theorem for two-point correlation
functions of the stress tensor.  It also gives a simple argument for a nonrenormalization theorem for
three-point functions which is much more difficult to demonstrate
directly and is only known to a few experts.
  It also 
suggests the $\theta$ independence of all Green's functions to all orders in the $1/N$ 
expansion of the Yang--Mills theory which agrees with the heuristic
argument  given in section 5.  It is undoubtedly true, and would be interesting
to demonstrate, that all of
these  results follow from
supersymmetry considerations in the conformally invariant
  Yang--Mills theory which are the image of  
 the powerful supersymmetry
constraints in IIB supergravity.

In addition to these nonrenormalization conditions we have also 
obtained expressions  for various
four-point 
correlation functions
in the theory.  These expressions are nonperturbative in $\gy$ and exact
through  next to
leading order in an expansion in $(\gy^2 N)^{-\ha}$.
   While we have not displayed these expressions in
detail in the Yang--Mills theory, they can easily be translated 
from the corresponding amplitudes in IIB supergravity  
in an \AdS5s5\ background.  This  shows that
superconformally invariant gauge theories contain exactly calculable
but 
highly nontrivial
correlation functions analogous to the Seiberg-Witten formulae of
nonconformally invariant
$\calN =2$ theories (which become  trivial in the conformally invariant
case).   It would again 
be interesting to find derivations of these terms from superconformal
Ward identities.

Of course, our nonperturbative string theoretic information is rather
limited.
It  covers only certain terms which are `protected' by supersymmetry in
the sense that they are given by integration over half the superspace.
Also, the mapping into the gauge theory is restricted to the strong coupling
expansion (large $\gy^2 N$) which corresponds to the $\alphaprime$ expansion of string theory.  
In order to understand the weak coupling regime (small $\gy^2 N$) it
will be necessary to understand exact properties of  the string theory
S-matrix, which is somewhat more difficult.

\vskip 0.6cm
\noindent{\bf Acknowledgements}:  We wish to acknowledge useful conversations with 
 Steven Shenker, Hirosi Ooguri, Gary Horowitz and other participants of 
the workshop on \lq Dualities in 
String Theory' at the Institute for Theoretical Physics, Santa Barbara, as well as
with Gary Gibbons, Hugh Osborn,  Dan Freedman, Igor Klebanov and  Ed
Witten.  We are also grateful to Lubos Motl for pointing out various
typographical errors in the first version of this paper.

 \listrefs

 \end